\documentclass[11pt]{article}

\usepackage{graphicx}
\usepackage{natbib}
 \usepackage{mathptmx}      
\usepackage{amsmath}
\usepackage{amssymb}
\usepackage{amsfonts}
\usepackage{lmodern}
\begin{document}

\title{Model Assisted Probability of Detection curves: New statistical tools and progressive methodology}




\author{Lo\"ic Le Gratiet \and Bertrand Iooss \and G\'eraud Blatman \and Thomas Browne \and Sara Cordeiro \and Benjamin Goursaud \and (EDF, France)}


\maketitle

\begin{abstract}
The Probability Of Detection (POD) curve is a standard tool in several industries to evaluate the performance of Non Destructive Testing (NDT) procedures for the detection of harmful defects for the inspected structure. 
Due to new capabilities of NDT process numerical simulation, Model Assisted Probability of Detection (MAPOD) approaches have also been recently developed. 
In this paper, a generic and progressive MAPOD methodology is proposed.
Limits and assumptions of the classical methods are enlightened, while new metamodel-based methods are proposed.
They allow to access to relevant information based on sensitivity analysis of MAPOD inputs.
Applications are performed on  Eddy Current Non Destructive Examination numerical data.

\end{abstract}

\section{Introduction}\label{intro}

In several industries, the Probability Of Detection (POD) curve is a standard tool to evaluate the performance of Non Destructive Testing (NDT) procedures \citep{PODeniq2010,MIL2009,meycra14}. 
The goal is to assess the quantification of inspection capability for the detection of harmful flaws for the inspected structure. 
For instance, for the French company of electricity (EDF), the potentialities of this tool are studied in the context of the Eddy Current Non Destructive Examination in order to ensure integrity of steam generators tubes in nuclear power plants \citep{maurice}.

However, high costs of the implementation of experimental POD campaigns combined with continuous increase in the complexity of configuration make them sometimes unaffordable. 
To overcome this problem, it is possible to resort to numerical simulation of NDT process (see for example \citet{rupbla14} for ultrasonics and \citet{rosper13} for eddy-current).
This approach has been called MAPOD for ``Model Assisted Probability of Detection'' \citep{tho08} (see also \citet{meycra14} for a survey and \citet{cal12} for a synthetic overview). 

The determination of this ``numerical POD'' is based on a four-step approach:
\begin{enumerate}
\item Identify the set of parameters that significantly affect the NDT signal;
\item Attribute a specific probability distribution to each of these parameters (for instance from expert judgment);
\item Propagate the input parameters uncertainties through the NDT numerical model;
\item Build the POD curve from standard approaches like the so-called Berens method \citep{Berens1988}.
\end{enumerate}
In POD studies, two main models have been proposed: POD model for binary detection representation (using hit/miss data) and POD model for continuous response (using the values of the NDT signal).
We focus in this work on POD model for continuous response, arguing that model-based data contain quantitative and precise information on the signal values that will be better exploited with this approach.

As it totally relies on a probabilistic modeling of uncertain physical variables and their propagation through a model, the MAPOD approach can be directly related to the uncertainty management methodology in numerical simulation (see \citet{derdev08} and \citet{baudut16} for a general point of view, and \citet{domfeu12} for illustration in the NDT domain).
This methodology proposes a generic framework of modeling, calibrating, propagating and prioritizing uncertainty sources through a numerical model (or computer code).
Indeed, investigation of complex computer code experiments has remained an important challenge in all domain of science and technology, in order to make simulations as well as predictions, uncertainty analysis or sensitivity studies.
In this framework, the numerical model $G$ just writes
\begin{equation}\label{eq:model}
Y = G(X) = G(X_1,\ldots,X_d)\;,
\end{equation}
with $X \in \mathbb{R}^d$ the random input vector of dimension $d$ and $Y \in \mathbb{R}$ a scalar model output.

However, standard uncertainty treatment techniques require many model evaluations and a major algorithmic difficulty arises when the computer code under study is too time expensive to be directly used.
For instance, it happens for NDT models based on complex geometry modeling and finite-element solvers.
This problem has been identified in \citet{cal12} who distinguishes ``semi-analytical'' codes (fast to evaluate but based on simplified physics) and ``full numerical'' ones (physically realistic but cpu-time expensive) which are the models of interest in our work.
For cpu-time expensive models, one solution consists in replacing the numerical model by a mathematical approximation, called a response surface or a metamodel. 
Several statistical tools based on numerical design of experiments, uncertainty propagation efficient algorithms and metamodeling concepts will then be useful \citep{fanli06}.
They will be applied, in this paper, in the particular NDT case of a POD curve as a quantity of interest.

The physical system of interest, the numerical model parameterization and the design of numerical experiments are explained in the following section.
The third section introduces four POD curves determination methods: the classical Berens method, a binomial-Berens method and two methods (polynomial chaos and kriging) based on the metamodeling of model outputs. 
In the fourth section, sensitivity analysis tools are developed by using the metamodel-based approaches. 
A conclusion synthesizes the work with a progressive strategy for the MAPOD process, in addition to some prospects.

\section{The NDT system}

\begin{figure}[!ht]
  \centering
  \includegraphics[width=0.6\textwidth]{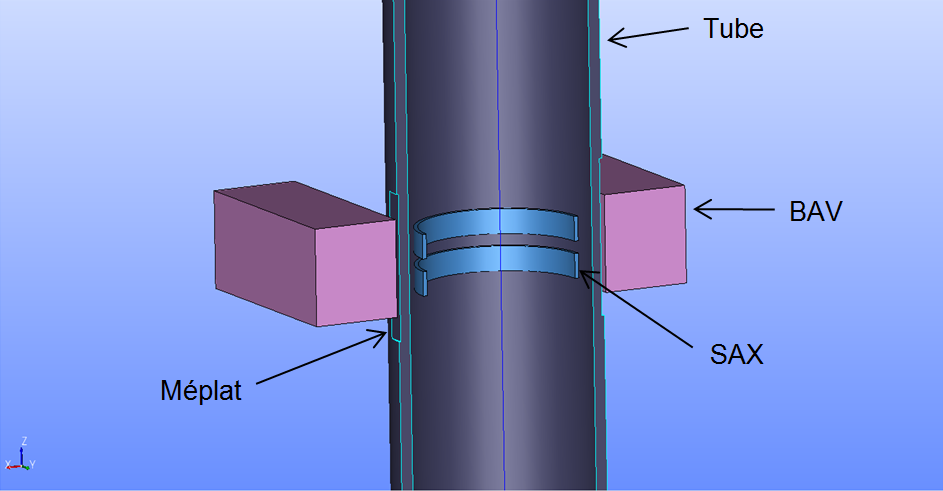}
\caption{Representation of the system under study (Tube, BAV and SAX).}
  \label{fig:shema}
\end{figure}

Our application case, shown in Figure \ref{fig:shema}, deals with the inspection by the SAX probe (an axial probe) of steam generator tubes to detect the wears, which are defects due to the rubbing of anti-vibration bars (BAV).
This configuration has been studied with the software Code\_Carmel3D (C3D) for several years.
This tool has demonstrated its ability to accurately simulate the signature of a wear with its influential parameters (mainly the BAV) \citep{SAXAVB}.

\subsection{The computer code and model parameterization}

\begin{figure}[!ht]
  \centering
  \includegraphics[width=0.6\textwidth]{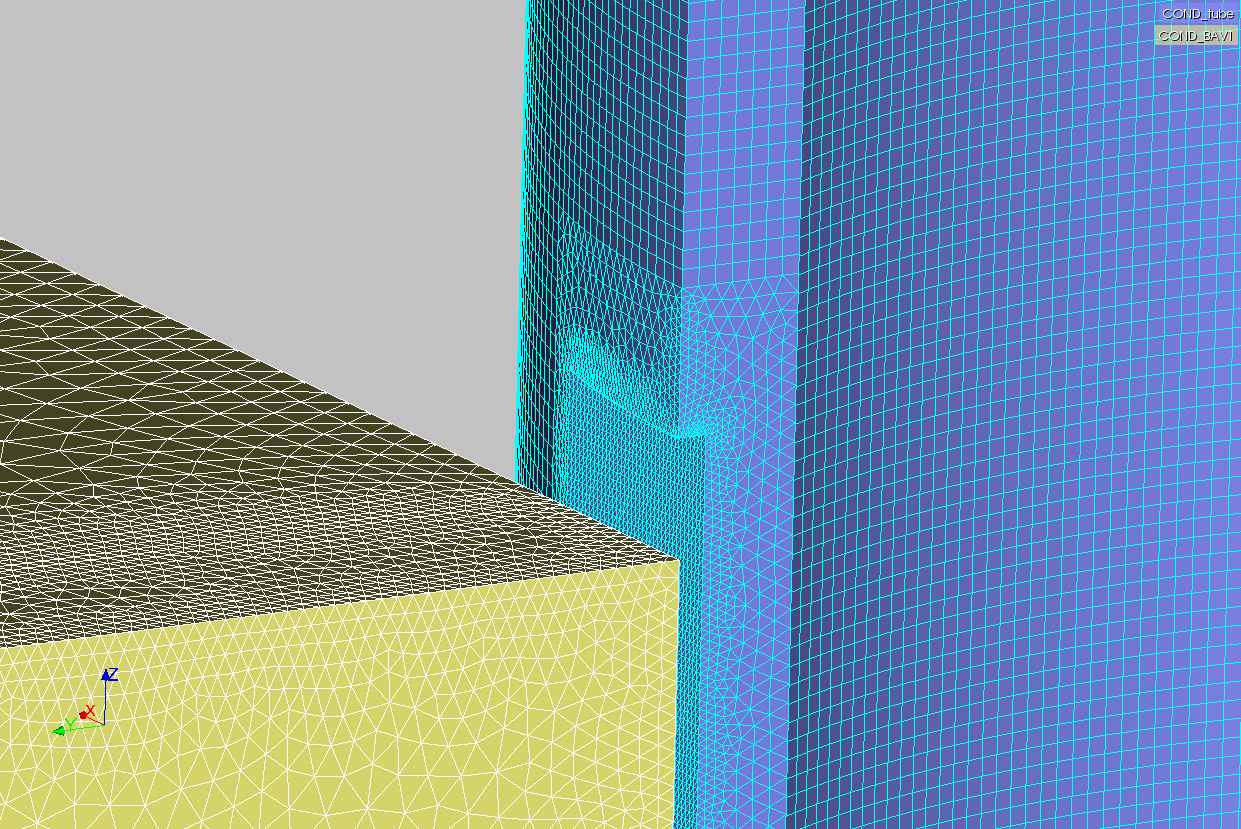}
\caption{Illustration of the mesh in the numerical model of NDT simulation.}
  \label{fig:mesh}
\end{figure}

The numerical simulations are performed by C3D, computer code derived from code\_Carmel developed by EDF R\&D and the L2EP laboratory of Lille (France).
This code uses the finite element method to solve the problem. Hence, there is a large flexibility for the parameters that can be taken into account (cf. Figure \ref{fig:mesh}).
The accuracy of the calculations can be ensured with a sufficiently refined mesh \citep{c3d}, using HPC capabilities if necessary. 

\begin{figure}[!ht]
  \centering
  \includegraphics[width=0.5\textwidth]{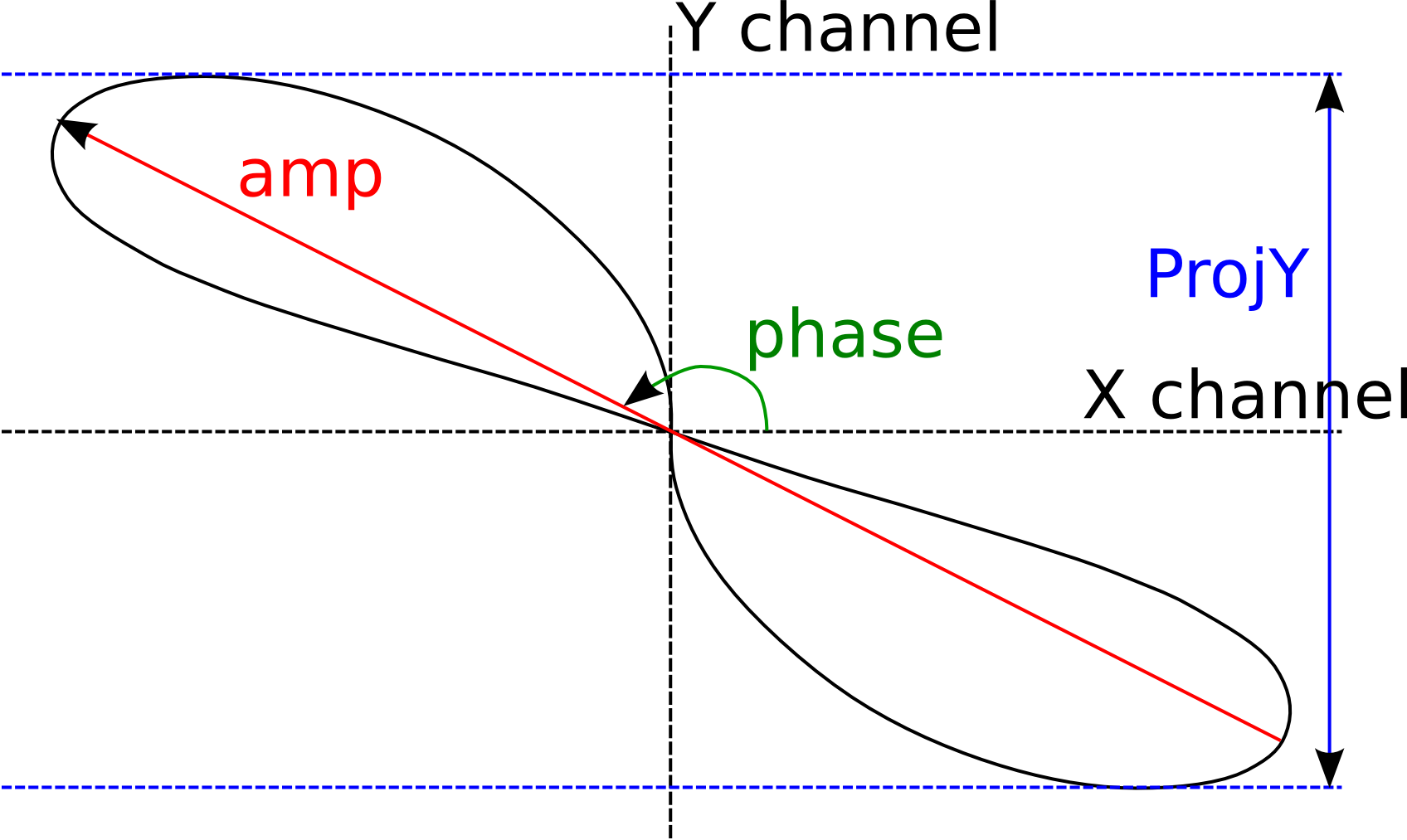}
\caption{Lissajous curve: output parameters of a NDT simulation for the SAX probe in differential mode (amp, phase and $ProjY$).}
  \label{fig:lissajous}
\end{figure}

The eddy-current non-destructive examinations are based on the change of the induction flux in the coils of the probe approaching a defect. 
When the tube is perfectly cylindrical, both coils of the probe get the same flux of induction.
If there is a defect, the flux are distinct and hence the differential flux, which is the difference between the flux in each coil, is non-zero: it is a complex quantity whose real part is the channel $X$ and the imaginary part is the channel $Y$.
Hence, when plotting the differential flux for each position of the probe, one gets a curve in the impedance plane, called a Lissajous curve.
The output parameters of a non-destructive examination are (as illustrated on Figure \ref{fig:lissajous}) :
\begin{itemize}
 \item the amplitude ($amp$), which is the largest distance between two points of the Lissajous curve,
 \item the phase, which is the angle between the abscissa axis and the line linking two points giving the amplitude, 
 \item the $Y$-projection ($ProjY$), which is the largest imaginary part of the difference between two points of the Lissajous curve.
\end{itemize}

\subsection{Input parameters and associated random distributions definition}\label{sec:inputs}

\begin{figure}[!ht]
  \centering
  \includegraphics[scale=0.7]{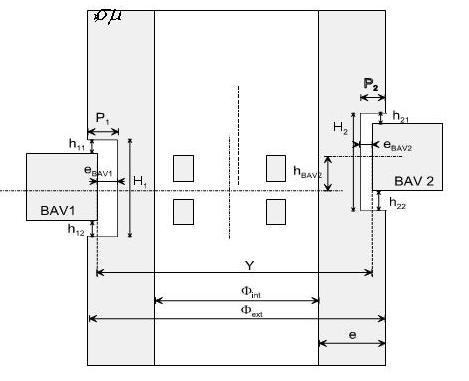}
\caption{Illustration of the considered inputs.}
  \label{fig:config}
\end{figure}

By relying on both expert reports and data simulations, the set of the input parameters which can have an impact on the code outputs have been defined.
Probabilistic models have also been proposed following deep discussions between NDT experts and statisticians. $\mathcal{N}(.,.)$ (resp. $\mathcal{U}[.,.]$) stands for Gaussian (resp. uniform) law.
These parameters are the following (see Fig. \ref{fig:config}): 
\begin{itemize}
\item $E \sim \mathcal{N}(a_E,b_E)$: pipe thickness (mm) based on data got from $5000$ pipes,
\item $h_1 \sim \mathcal{U}[a_{h_1},b_{h_1}]$: first flaw height (mm),
\item $h_2 \sim \mathcal{U}[a_{h_2},b_{h_2}]$: second flaw height (mm), 
\item $P_1 \sim \mathcal{U}[a_{P_1},b_{P_1}]$: first flaw depth (mm),
\item $P_2 \sim \mathcal{U}[a_{P_2},b_{P_2}]$: second flaw depth (mm),
\item $ebav_1 \sim \mathcal{U}[-P_1+a_{ebav_1},b_{ebav_1}]$: length of the gap between the BAV and the first flaw (mm),
\item $ebav_2 \sim \mathcal{U}[-P_2+a_{ebav_2},b_{ebav_2}]$: length of the gap between the BAV and the second flaw (mm). 
\end{itemize}
All these input parameters are synthesized in a single input random vector $(E,h_1,h_2,P_1,P_2,ebav_1,ebav_2)$.

As displayed in Figure \ref{fig:config}, we consider the occurrence of one flaw on each side of the pipe due to BAV. To take this eventuality into account in the computations, $50\%$ of the experiments are modeled with one flaw, and $50\%$ with two flaws.

\subsection{Definition of the design of numerical experiments}

In order to compute the output of interest with C3D, it is necessary to choose the points in the variation domain of the inputs (called the input set). 
This dataset, called ``design of experiments'', has to be defined at the very beginning of the study, which is to say before any numerical simulation. 
A classical method consists in building the design of experiments by randomly picking different points of the input set, obtaining a so-called Monte Carlo sample. 
However, a random sample can lead to a design which does not properly ``fill-in'' the input set \citep{fanli06}.
A better idea would be to spread the numerical simulations all over the input set, in order to avoid some empty big subsets. 

To this effect it is more relevant to choose the values according to a deterministic rule, such as a quasi-Monte Carlo method, for instance a Sobol' sequence. 
Indeed, for a size of design $N$, it is proved that this design often happens to be more precise than the standard Monte Carlo method \citep{fanli06}.
Given the available computing time (several hours per model run), a Sobol' sequence of size $100$ is created, and $100$ model outputs are obtained after the computer code ($G$) runs.

\section{Methods of POD curves estimation}

In this section several methods (from the simplest relying on strong assumptions to the most complex) are presented and applied.
The objective is to build the POD curve as a function of the main parameter of interest, related to the defect size.
As there are two defects in the system, $a:=\max(P_1,P_2)$ is chosen as the parameter of interest. 

\begin{figure}[!ht]
  \centering
	\includegraphics[scale=0.6]{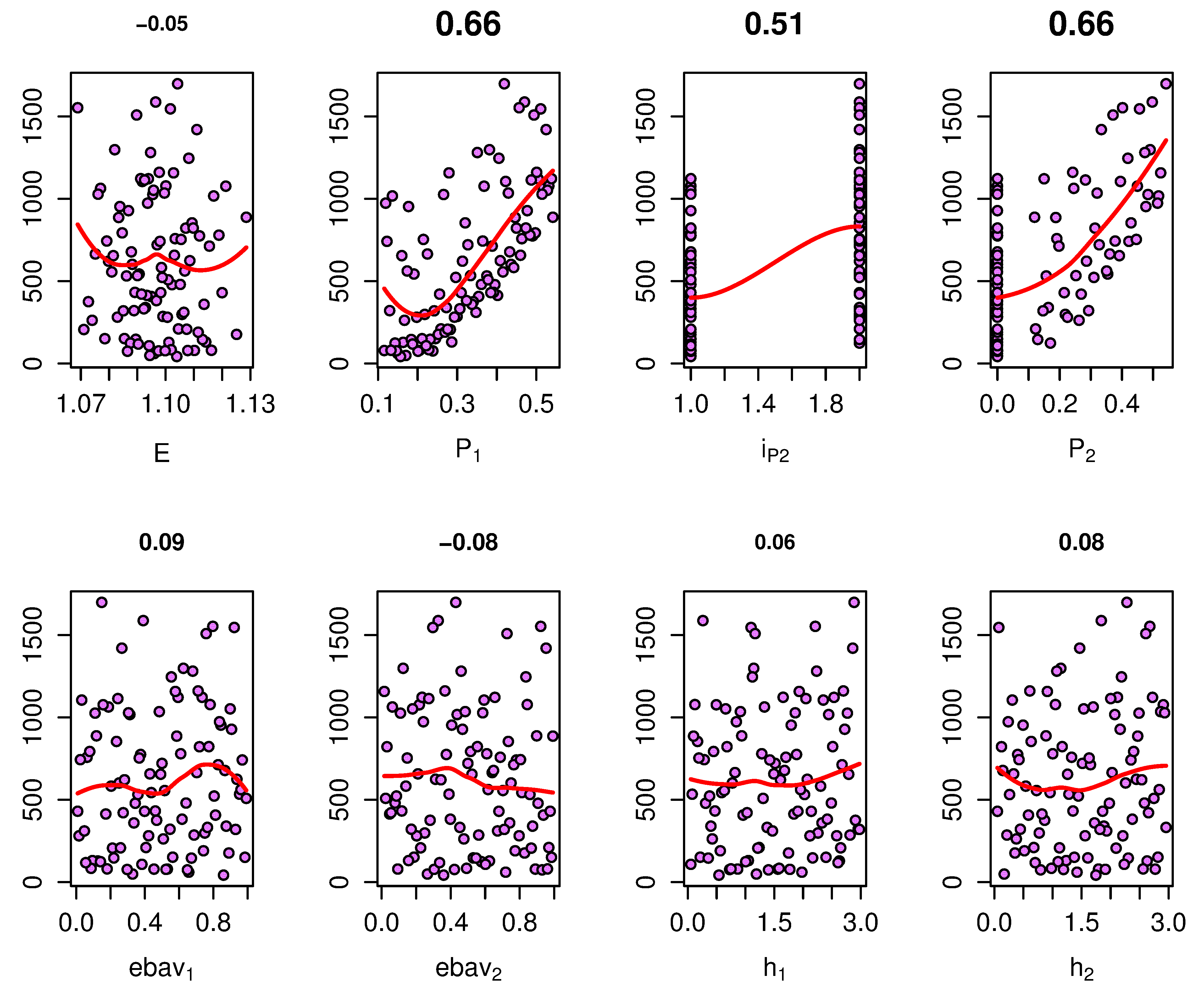}
\caption{$ProjY$ with respect to the input parameters. On each plot, the solid curve is a local polynomial smoother and the upper number is the corresponding correlation coefficient between the input (in abscissa) and $ProjY$ (in ordinate).}\label{fig:ProjY}
\end{figure}

By using the computer code C3D, one focuses on the output $ProjY$ which is a projection of the simulated signal we would get after NDT process. 
The other inputs are seen as random variables, which makes $ProjY$ itself an other random variable. 
The model (\ref{eq:model}) writes now
\begin{equation}\label{eq:ProjY}
ProjY = G(a,X)
\end{equation}
with the random vector $X=(E,h_1,h_2,ebav_1,ebav_2)$.
The effects of all the input parameters $(a,X)$ are displayed in Figure \ref{fig:ProjY}.
The bold values are the correlation coefficients between the output $ProjY$ and the corresponding input parameter.
Strong influences of $P_1$ and $P_2$ on $ProjY$ are detected.
$i_{P2}$ is the binary variable governing the presence of one flaw ($i_{P2}=1$) or two flaws ($i_{P2}=2$).

Given a threshold $s>0$, a flaw is considered to be detected if $ProjY>s$. Therefore the one dimensional POD curve is denoted by: 
\begin{equation}\label{eq:POD}
 \forall a>0 \quad \mbox{POD}(a)= \mathbb{P}\left( G(a,X) > s \mid a \right) \;.
\end{equation}
Four different regression models of $ProjY$ are proposed in the following, in order to build an estimation of the POD curve. 
Numerical simulations are computed for the $N=100$ points of the design of experiments.

\subsection{Data linearization step}

\begin{figure}[!ht]
  \centering
	\includegraphics[width=0.49\textwidth]{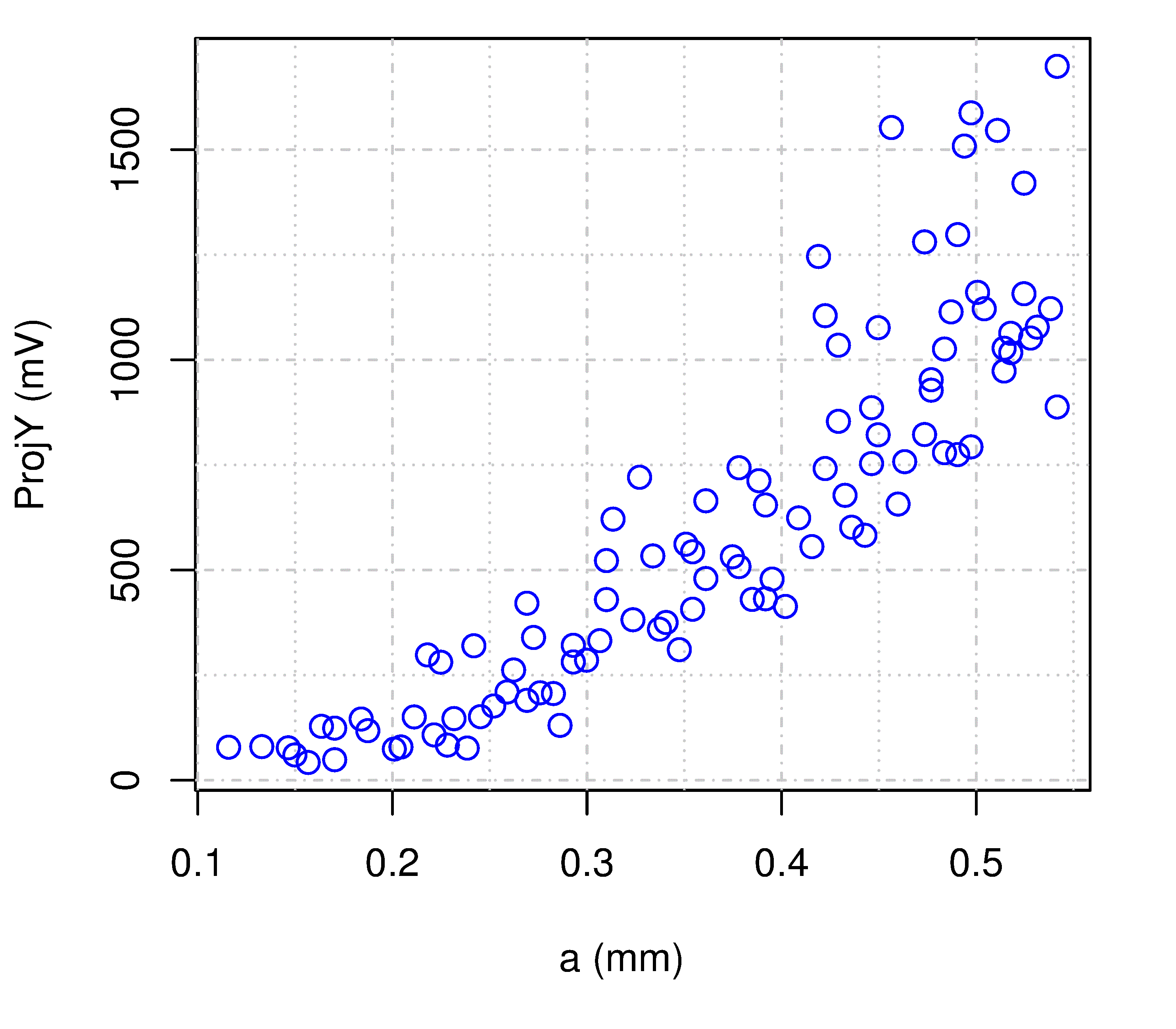}
	\includegraphics[width=0.49\textwidth]{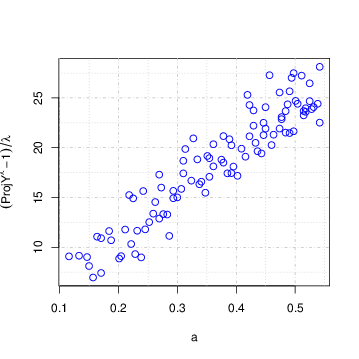}
\caption{Model response with respect to $a$. Left: Initial data ($ProjY$ as response); Right: Linearized data ($y_{ProjY}$ as response) by Box-Cox transformation with parameter $\lambda = 0.3$ of the response $ProjY$.}
  \label{fig:boxcoxAvsa}
\end{figure}

All the POD methods consist in a  (linear or non-linear) regression of the output $ProjY$. 
Then, a data linearization is useful to improve the adequacy of the models.
This can be made by a Box-Cox transformation \citep{BoxCox1964} of the output, which means that we now focus on: 
\begin{equation}
y = \frac{ProjY^{\lambda}-1}{\lambda}\;.
\end{equation}
$\lambda$ is determined by maximum likelihood as the real number that offers the finest linear regression of $y$ regarding the parameter $a$ (see Figure \ref{fig:boxcoxAvsa}).
The same transformation has to be applied to the detection threshold $s$.
In the following, we keep $s$ for the notation of this threshold.
It is important to note that this transformation is useful for all the different POD methods \citep{domfeu12}.

\subsection{Berens method \citep{Berens1988}}\label{sec:Berens}

\begin{figure}[!ht]
  \centering
	\includegraphics[scale=0.5]{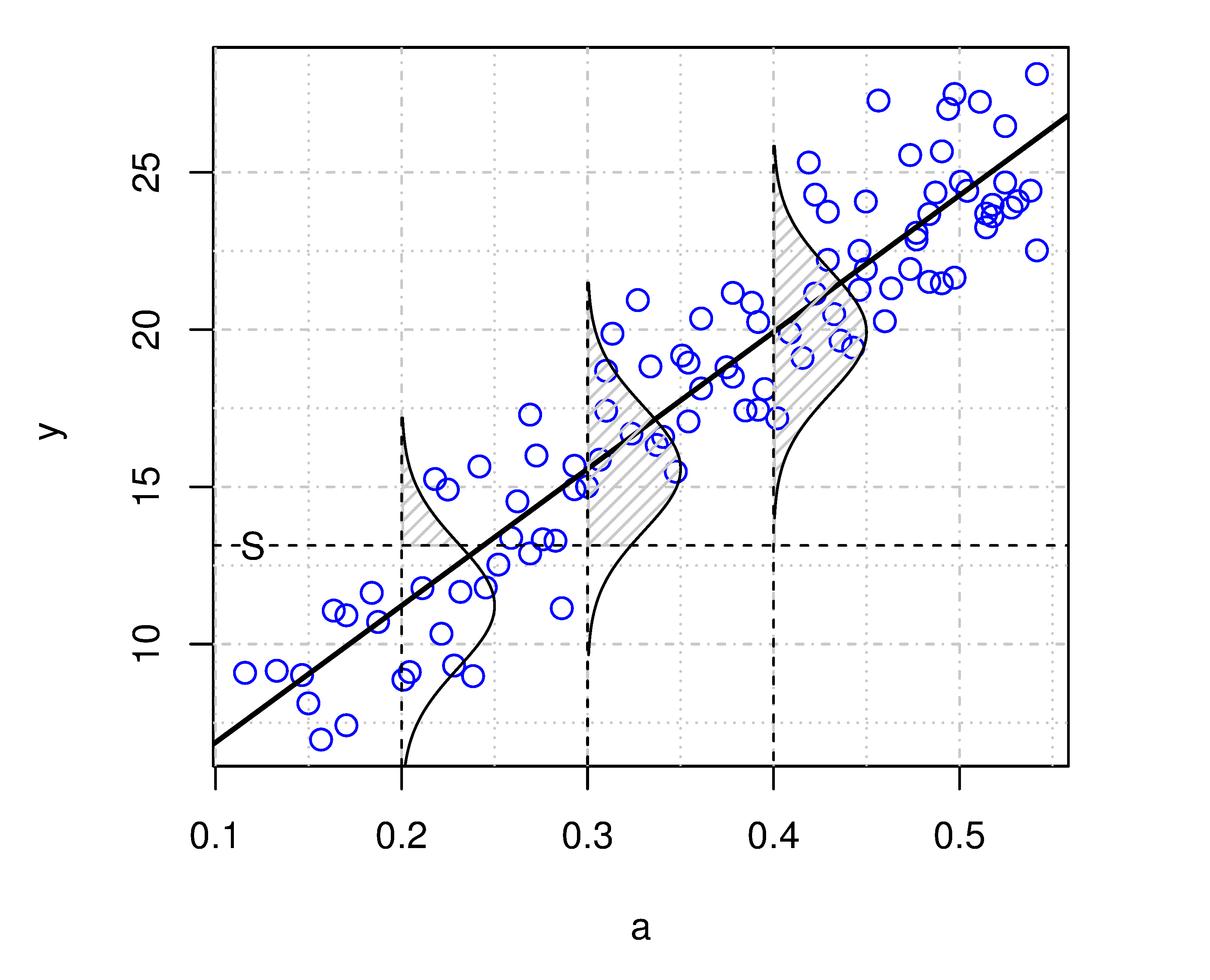}
\caption{Linear model illustration. The Gaussian predictive distributions for  $a = \max\left( P_1, P_2 \right)=0.2$, $0.3$ and $0.4$ are given. The horizontal line represents the detection threshold $s$.}
  \label{fig:linearmodel}
\end{figure}

The Berens model, based on $y$, is defined as
\begin{equation}\label{eq:berens}
y(a)=\beta_0 + \beta_1a + \epsilon,
\end{equation}
 with $\epsilon$ the model error such as $\epsilon \sim \mathcal{N}\left(0,\sigma_{\epsilon}^2 \right)$. Maximum likelihood method provides the estimators $\hat{\beta_0}, \hat{\beta_1} \; \text{and} \; \hat{\sigma_{\epsilon}}$. Hence the model implies the following result: $\forall a>0, \quad y(a) \sim \mathcal{N}\left(\hat{\beta_0} +\hat{\beta_1}a, \hat{\sigma_{\epsilon}}^2 \right)$. 
On our data, we obtain $\hat{\beta_0}=2.52$, $\hat{\beta_1}=43.48$ and $\hat{\sigma}_\epsilon=1.95$, which leads to the linear model represented in Figure \ref{fig:linearmodel}.

\begin{figure}[!ht]
  \centering
	\includegraphics[scale=0.5]{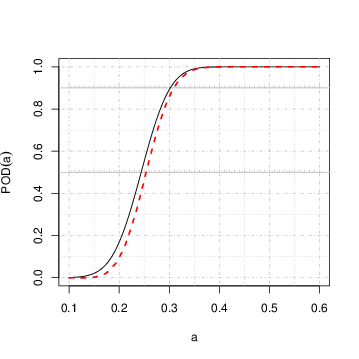}
\caption{Results of Berens method: POD curve estimation (solid curve) and POD lower curve  (dashed curve) of the POD $95\%$-confidence interval.}
  \label{fig:PODgaussconfident}
\end{figure}

With the normality hypothesis, as displayed in Figure \ref{fig:linearmodel}, the values of the POD curve can be easily estimated, giving the POD curve of Figure \ref{fig:PODgaussconfident}.
By considering the error that is provided by the property of a maximum likelihood estimator in a case of a linear regression, we can use this uncertainty on both $\beta_0$ and $\beta_1$ to build non-asymptotic confidence intervals.  
Indeed, the Gaussian hypothesis on $\epsilon$ makes it possible to obtain the prediction law of $\beta_0$ and $\beta_1$ conditionally to $\sigma^2_\epsilon$ :
\begin{equation}
 {\beta_0 \choose \beta_1} \sim  \mathcal{N}\left( \hat{\beta} = {\hat{\beta_0} \choose \hat{\beta_1}} ,\sigma^2_\epsilon \left(\bf{X}^T \bf{X}\right)^{-1} \right),
\end{equation}
with $\bf{X}$ the data input matrix:
\[
\bf{X} = \left(\begin{array}{rl}
1 & a_1 \\
1 & a_2 \\
\vdots & \vdots \\
1 & a_N
\end{array}\right).
\]

Classical results on linear regression theory state that the variance $\sigma^2_\epsilon$ follows a  chi-2 distribution with $N-2$ degrees of freedom:
\begin{equation}
\frac{(N-2) \hat{\sigma_\epsilon}^2}{\sigma_\epsilon^2} \sim \chi^2_{N-2},
\end{equation}
where 
\begin{equation}
\hat{\sigma_\epsilon}^2 = \frac{ \left( y^N - \bf{X} \hat{\beta}\right)^T\left( y^N - \bf{X} \hat{\beta}\right)}{N-2},
\end{equation}
with $y^N=(y(a_1),\ldots,y(a_N))$ the data output sample.
Then, we can obtain a sample $(\beta_{0},\beta_{1},\sigma_\epsilon^2)$ by simulating $\sigma_\epsilon^2$ then $\beta_0 \choose \beta_1$ conditionally to $\sigma_\epsilon^2$. 
From this sample, we get a sample of $\mbox{POD}(a)$ via the formula:
\begin{equation}
1 - \Phi \left( (s - \beta_0 - \beta_1 a)/{\sigma_\epsilon} \right),
\end{equation}
where $\Phi$ is the standard Gaussian distribution.
By simulating a large number of POD samples, we can deduce some confidence intervals.
The 95\%-confidence lower bound of the POD curve is illustrated in Figure \ref{fig:PODgaussconfident}.

From the estimated POD of Figure \ref{fig:PODgaussconfident}, we obtain $a_{90} \simeq 0.30$ mm for the defect size detectable with a $90\%$-probability.
Taking into account the confidence interval, we obtain $a_{90/95} \simeq 0.31$ mm for the minimal defect size detectable with a $90\%$-probability with a $95\%$-confidence.
This means that the defect size that we detect in $90\%$ of cases has a $95\%$-probability to be smaller than $0.31$ mm.

In conclusion, we remind that the Berens method stands on two hypotheses that have to be validated:
\begin{itemize}
\item the linearity relation between $y$ and $a$ (after the Box-Cox transformation) that can be studied via classical linear regression residuals analysis \citep{chr90}.
On our data, we have for instance $R^2=88\%$ for the regression coefficient of determination, indicator which denotes the explained variance of the linear regression;
\item the Gaussian distribution, homoscedasticity and independence of the residuals that can be studied via many statistical tests (see for instance \citet{walpro97}).
On our data, we have the following p-values: $0.62$ for Kolmogorov-Smirnov test (Gaussian distribution), $0.10$ for Anderson-Darling test (Gaussian distribution), $0.82$ fo Breusch-Pagan test (homoscedasticity) and $0.12$ for Durbin-Watson test (non correlation).
We conclude that,  with a $90\%$-confidence, the homoscedasticity and non-correlation hypotheses of $\epsilon$ cannot be rejected, but the normality hypothesis of $\epsilon$ can be rejected. 
\end{itemize}

\subsection{Binomial-Berens method}

\begin{figure}[!ht]
  \centering
	\includegraphics[scale=0.6]{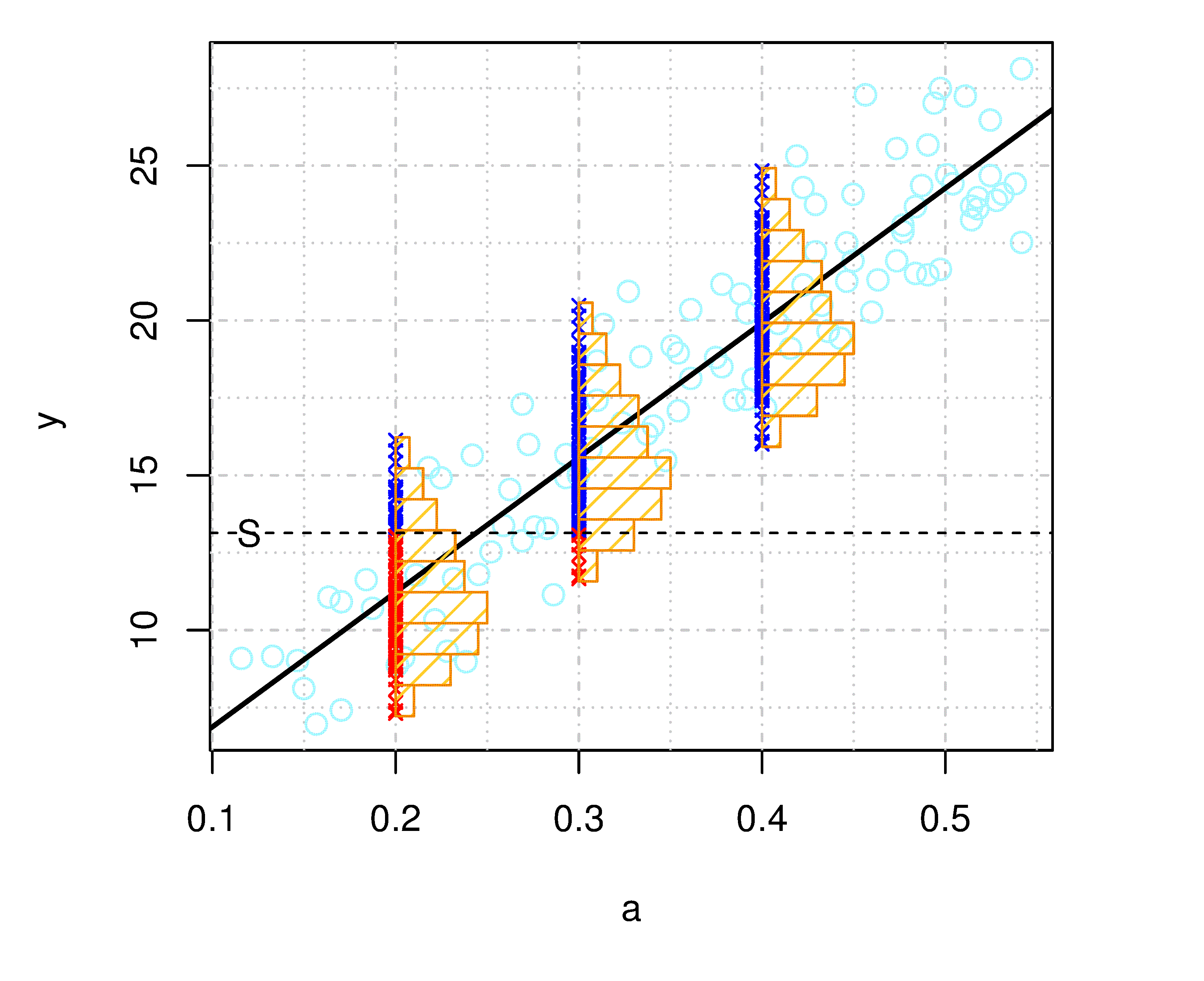}
\caption{Binomial-Berens method: Berens method without normal hypothesis. The Gaussian densities are replaced by the sample histogram. The horizontal line represents the detection threshold $s$.}
  \label{fig:binomialmodel}
\end{figure}

Here we keep the linear regression on $y$, which is: $\forall a >0 \quad y= \hat{\beta_0}+\hat{\beta_1}a+ \epsilon$ but we do not assume that $\epsilon$ is Gaussian anymore. However the errors are still assumed to be independent and identically distributed. 
We then consider that we have $N$ of its realizations which we regroup in the following vector
\begin{equation}
\epsilon^N=y^N-\hat{\beta_0}-\hat{\beta_1}a^N.
\end{equation}
Therefore we build its histogram and we add it to the prediction of the linear model as shown in Figure \ref{fig:binomialmodel}. 
By using the i.i.d. property of $\epsilon$, let us consider that we have $N$ realizations of the random value $y(a)$ for $a>0$.
We propose to use them to estimate the probability for $y(a)$ to exceed the threshold $s$ (see Figure \ref{fig:binomialmodel}).

For each $a>0$, let $N_s(a)$ be the number of realizations of the random variable $y(a)$ that are higher than $s$. That is to say: 
\begin{equation}
N_s(a) =\text{Card} \left( \left\lbrace \left(\epsilon_i \right)_{i \in \{1,...,N\} } \mid \hat{\beta_0}+\hat{\beta_1}a+ \epsilon_i >s \right\rbrace \right) .
\end{equation}
Therefore an estimation of $\mbox{POD}(a)$ is given by $\frac{N_s(a)}{N}$, with $N_s(a) \sim \mathcal{B}\left(N,\mbox{POD}(a)\right)$, with $\mathcal{B}$ the binomial probability law. 
The assumption on $N_s(a)$ distribution can then be used to build confidence intervals on the value of $\mbox{POD}(a)$, for $a>0$. 

Let us note that the Binomial-Berens method only requires to validate the linear relation between $y$ and $a$.
For the $90\%$-level defect, we obtain $a_{90} \simeq 0.30$ mm and $a_{90/95} \simeq 0.305$ mm.
A slight difference with the classical Berens method is present for $a_{90/95}$.

\subsection{Polynomial chaos method}

As some criticism could be made at some point regarding the simplistic linear model of equation (\ref{eq:berens}), let us build a metamodel \citep{fanli06} of the transformed output $y$. 
Now the influence of the other inputs (described in Section~\ref{sec:inputs}) are explicitly mentioned in the model whereas it used to be all included in $\epsilon$. 
The model response of interest, e.g. the $Y$-projection, is represented as a ``pure'' function of $X$ (i.e. without additional noise):
\begin{equation} 
Y = G(a,X) . 
\end{equation} 
The so-called polynomial chaos (PC) method \citep{Soize2004, Blatman_JCP_2011} consists in approximating the response 
onto a specific basis made of orthonormal polynomials:
\begin{equation} 
Y \; \; \approx \; \; \widehat{Y} = \sum_{j=0}^{P-1} a_j \psi_j(a,X) \; ,
\end{equation} 
where the $\psi_j$'s are the basis polynomials and the $a_j$'s are deterministic coefficients 
which fully characterize the model response and which have to be estimated. The 
orthonormality property reads:
\begin{equation} 
\mathbb{E}\left[\psi_i(a,X)\psi_j(a,X)\right] \; \; = \; \; 
1 \mbox{  if  } i=j \mbox{  else  } 0 \; .
\end{equation} 
The derivation of sensitivity indices (see Section~\ref{sec:Sobol}) of the response is direclty obtained by simple algebraic 
operations on the coefficients $a_j$. The latter are computed based on the 
experimental design and the associated model evaluations by least squares. 

PC approximations are computed with several values for the total degree, and their accuracies are compared in terms of predictivity coefficient $Q^2$, itself based on the leave-one-out error. 
The greatest accuracy is obtained with a linear 
approximation (i.e. with degree equal to one), with $Q^2=88\%$. This PC 
representation reads:
\begin{equation} \label{eq_PCE}
    \begin{split}
    \widehat{Y} \; \; \simeq \; \;  27.9 \; &- \; 0.5 \; \psi_{1}(E) \; + \; 
    11.4 \; \psi_2(a) \; + \; 0.7 \; \psi_3(ebav_1) \\& \; + \; 0.3 \; \psi_3(ebav_1) \; + \; 
    0.4 \; \psi_4(h_1) \; + \; 1.0 \; \psi_5(h_2) \quad \mbox{ (mV)}
    \end{split}
\end{equation}

As in the Berens model in Section~\ref{sec:Berens}, it is assumed that the 
approximation error is a normal random variable $\epsilon$ with zero mean and 
standard deviation equal to $\sigma_\epsilon$, that is:
\begin{equation}
    Y \; \; \approx \; \; \widehat{Y}  + \epsilon , \quad \epsilon \sim \mathcal{N}(0,\sigma_\epsilon^2)
\end{equation}
Thus the POD associated with a given defect size $a$ can be approximated by:
\begin{equation} \label{eq_pod_chaos}
    \mbox{POD}(a) \; = \; \mathbb{P}( Y > s \mid a) \; = \;
    \mathbb{P}\left(\widehat{Y}(a,X) + \epsilon > s\right).
\end{equation}
 For any value of $a$, this probability is estimated 
by Monte Carlo simulation of the random quantities $X$ and 
$\epsilon$ ($10^4$ random values are drawn). 

\begin{figure}[!ht]
  \centering
	\includegraphics[scale=0.4]{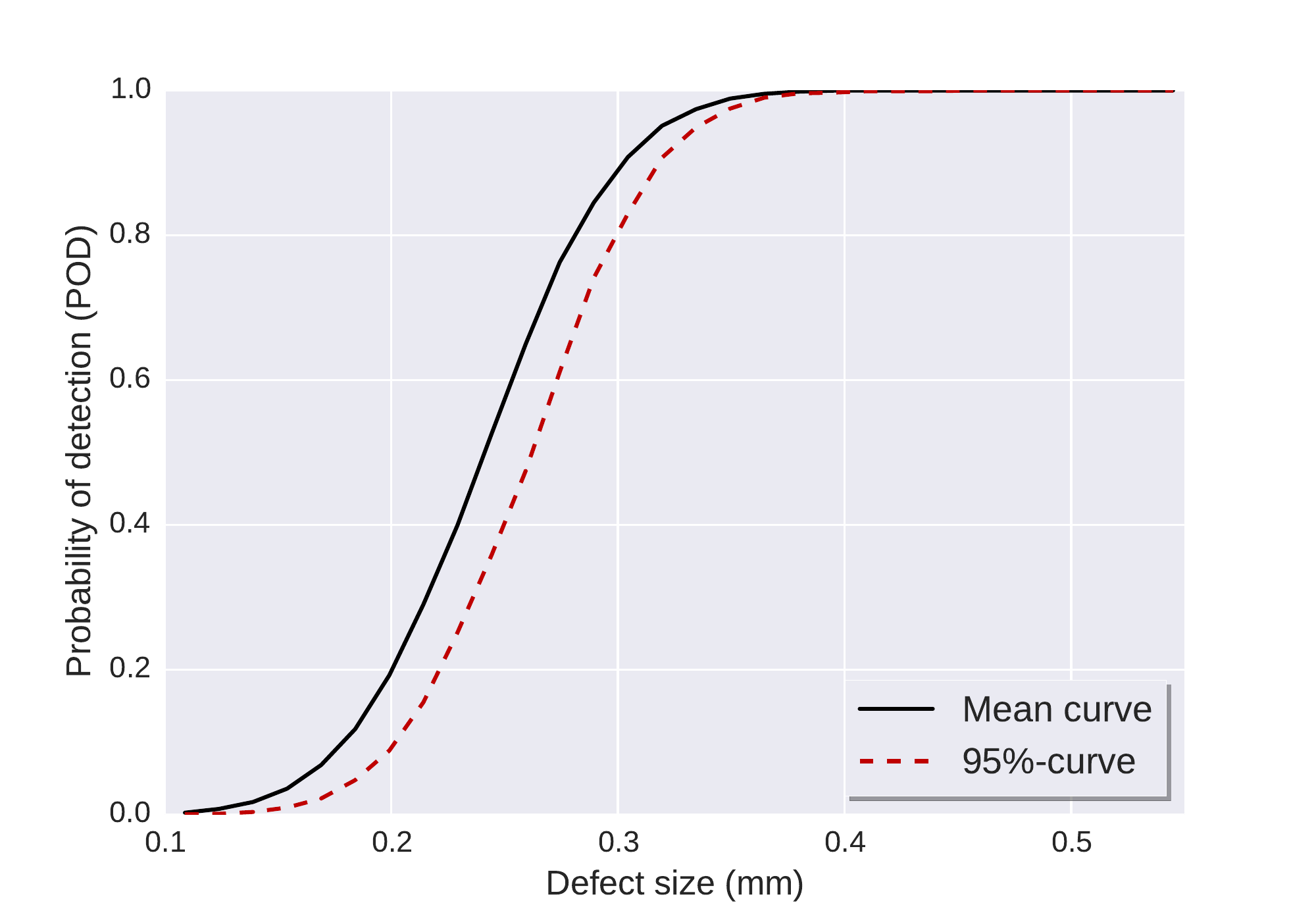}
\caption{Average and $95\%$-POD curves based on a PC approximation}
  \label{fig:chaospod}
\end{figure}

Note that this estimate 
relies upon the assumption that the chaos coefficients are perfectly calculated. 
However, their estimation is affected by uncertainty due to the approximation 
error ($1-Q^2=12\%$ of unexplained variance of the $Y$-projection) and the limited 
number of available evaluations of C3D. 
As for the Berens model, 
standard theorems related to linear regression hold for the PC expansions and 
can be used to define the probability distribution of the chaos coefficients 
and the residual standard deviation $\sigma_\epsilon$. Based on these 
results, $150$ sets of both quantities are randomly generated and each 
realization is used to compute the POD (Eq.(\ref{eq_pod_chaos})). Hence, for 
any $a$, a sample of $150$ values of $\mbox{POD}(a)$ is obtained. We computed its 
$5\%$-empirical quantile in order to construct the $95\%$-POD curve. The average 
and the $95\%$-POD curves are plotted in Figure~\ref{fig:chaospod}. 
The characteristic defect sizes (defined in the previous sections) are 
given by $a_{90} \simeq 0.30 \: \mbox{mm}$ and $a_{90/95} \simeq 0.32 \: \mbox{mm}$. 

It has to be noted that the chaos results are closed to the ones obtained by the 
Berens approach. Indeed, the PC representation~(\ref{eq_PCE}) is similar to the 
Berens model (\ref{eq:berens}) as all the coefficients except the mean value 
and the factor related to $a$ are relatively insignificant in our application case. 
Furthermore, 
it is also supposed that the residuals are independent realizations of a normal 
random variable. As discussed previously, this assumption can be rejected by 
statistical tests. 
Another kind of metamodel, namely kriging, is based on the weaker and more realistic assumption of correlated normal residuals (the correlation between two model evaluations increases as the related inputs get closer). 
This is the scope of the next section.

\subsection{Kriging method}

We turn now to a probabilistic metamodel technique, which is the Gaussian process regression \citep{sacwel89}, first proposed by \citet{Demeyer} for POD estimation.
Since the linear trend used in the Berens method was rather relevant, we keep it as the mean of the Gaussian process that we are about to use. 
The kriging model is defined as follows: 
\begin{equation} 
Y(a,X) = \beta_0 + \beta_1 a + Z(a,X) , 
\end{equation}
where $Z$ is a centered Gaussian process. We make the assumption that $Z$ is second order  stationary with variance $\sigma^2$ and covariance Mat\'ern 5/2 parameterized by its lengthscale $\theta$ ($\theta \in \mathbb{R}^6$ in our application case). 
Thanks to the maximum likelihood method, we can estimate the values of the so far-unknown parameters: $\beta_0, \beta_1, \sigma^2$ and $\theta$ (see for instance \citet{marioo08} for more details).

Kriging provides an estimator of $Y(a,X)$ which is called the kriging predictor and written $\widehat{Y_{P}}(a,X)$.
On our data, we compute the predictivity coefficient $Q^2$ in order to quantify the prediction capabilities of this metamodel \citep{marioo08}.
We obtain $Q^2=90\%$.

In addition to the kriging predictor, the kriging variance $\sigma_Y^2(a,X)$ quantifies the uncertainty induced by estimating $Y_{P}(a,X)$ with $\widehat{Y_{P}}(a,X)$. 
Thus, we have the following predictive distribution:
\begin{equation}\label{eq:kriging}
\forall x \quad \left(Y(a,X) \mid y^N \right) \sim   \mathcal{N} \left( \widehat{Y_{P}}(a,X), \sigma_Y^2(a,X) \right) 
\end{equation}
where $\widehat{Y_{P}}(a,X)$ (the kriging mean) and $\sigma_Y^2(a,X)$ (the kriging variance) can both be explicitly estimated. 

Obtaining the POD curve consists in replacing $Y = G(a,X)$ by its kriging metamodel (\ref{eq:kriging}) in (\ref{eq:POD}).
 Hence we can estimate the value of $\mbox{POD}(a)$, for $a>0$ from:
\begin{equation}\label{eq:PODkriging}
\mbox{POD}(a) = \mathbb{P}\left((Y(a,X) \mid y^N) > s | a \right) .
\end{equation}
Two sources of uncertainty have to be taken into account in (\ref{eq:PODkriging}): the first coming from the parameter $X$ and the second coming from the Gaussian distribution in (\ref{eq:kriging}).
From (\ref{eq:PODkriging}), the following estimate for $\mbox{POD}(a)$ can be deduced:
\begin{equation} 
\mbox{POD}(a) = \mathbb{E}_X \left[ 1 - \Phi\left(\frac{s-\widehat{Y_{P}}(a,X)}{\sigma_Y(a,X)}\right)\right].
\end{equation}
This expectation is estimated using a classical Monte Carlo integration procedure.

\begin{figure}[!ht]
  \centering
	\includegraphics[scale=0.4]{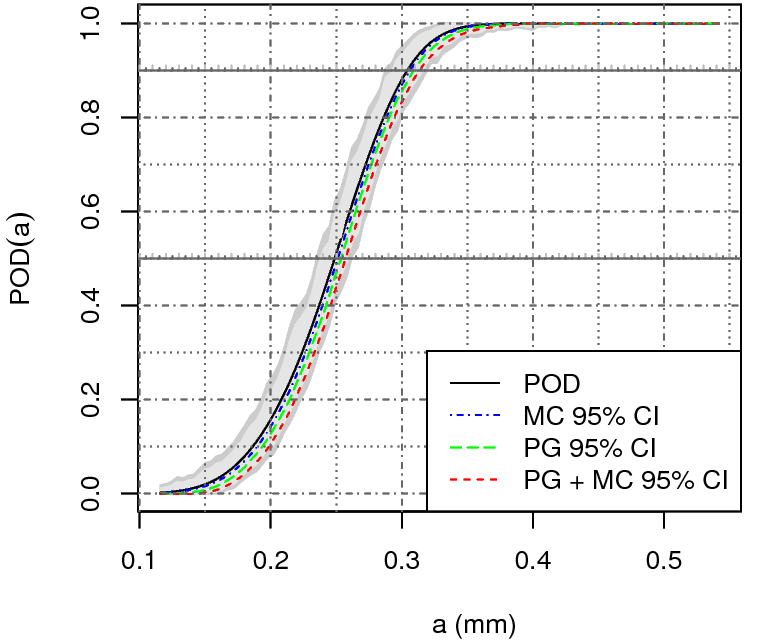}
\caption{Example of POD curves estimated with a kriging model.}
  \label{fig:krigingpod}
\end{figure}

By using the uncertainty implied by the Gaussian distribution regressions, one can build new confidence intervals as it is illustrated in Figure \ref{fig:krigingpod}.
We visualize the confidence interval induced by the Monte Carlo (MC) estimation, the one induced by the kriging (PG) approximation and the total confidence interval (including both approximations: PG+MC).
For the $90\%$-level defect, we obtain $a_{90} \simeq 0.305$ mm and $a_{90/95} \simeq 0.315$ mm.

The four methods discussed in this section have given somewhat similar results.
This will be discussed in the conclusion of this paper, which also introduces a general and methodological point of view for the numerical POD determination.

\section{Sensitivity analysis on POD curve}

Sensitivity analysis allows to determine those parameters that mostly influence on model
response.
In particular, global  sensitivity analysis methods (see \citet{ioolem15} for a recent review) take into account the overall uncertainty ranges of the model input parameters.
In this section, we propose new global sensitivity indices attached to the whole POD curve.
We focus on the variance-based sensitivity indices, also called Sobol' indices, which are the most popular tools and were proved robust, interpretable and efficient.

\subsection{Sobol' indices on scalar model output}\label{sec:Sobol}

If all its inputs are independent and $\mathbb{E}(Y^2)<\infty$, the variance of the numerical model $Y=G(X_1,\ldots,X_d)$ can be decomposed in the following sum:
\begin{equation}
\mbox{Var}(Y) = V = \sum_{i=1}^d V_i + \sum_{i<j} V_{ij} + \ldots + V_{1\ldots d} \,
\end{equation}
with $V_i=\mbox{Var}[\mathbb{E}(Y|X_i)]$, $V_{ij}=\mbox{Var}[\mathbb{E}(Y|X_i X_j)]-V_i - V_j$, etc.
Then, $ \forall i,j=1\ldots d, \;i<j$, the Sobol' indices of $X_i$ write \citep{sob93}:
\begin{equation}
S_i = \frac{V_i}{V} \;, S_{ij} = \frac{V_{ij}}{V} \;, \ldots, \mbox{ and } T_i = S_i+S_{ij}+\ldots \;.
\end{equation}
The first-order Sobol' index $S_i$ measures the individual effect of the input $X_i$ on the variance of the output $Y$, while the total Sobol' index $T_i$ measures the $X_i$ effect and all the interaction effects between $X_i$ and the other inputs (as the second-order effect $S_{ij}$).
$T_i$ can be rewritten as $T_i=\displaystyle 1 - \frac{V_{-i}}{V}$ with $V_{-i}=\mbox{Var}[\mathbb{E}(Y|X_{-i})]$ and $X_{-i}$ the vector of all inputs except $X_i$.

These indices are interpreted in terms of percentage of influence of the different inputs on the model output uncertainty (measured by its variance).
They have been proven to be useful in many engineering studies involving numerical simulation models \citep{derdev08}.

\subsection{Sobol' indices on POD}\label{sec:SobolPOD}

In order to define similar sensitivity indices for the whole POD curve, we first define the following quantities:
\begin{equation}\label{eq:PODX}
\begin{array}{rcl}
\mbox{POD}_X(a) &= &\mathbb{P}( Y > s \mid a,X) \;,\\
\mbox{POD}_{X_i}(a) &= &\mathbb{P}( Y > s \mid a,X_i) \;,\\
\mbox{POD}_{X_{-i}}(a) &= &\mathbb{P}( Y > s \mid a,X_{-i}) \;,\\
D & = &\mathbb{E}\|\mbox{POD}(a)-\mbox{POD}_X(a)\|^2
\end{array}
\end{equation}
with $\|.\|$ the euclidean norm.
The POD Sobol' indices are then defined by:
\begin{equation}
\begin{array}{rcl}
S_i^{\mbox{\tiny POD}} & = & \displaystyle \frac{\mathbb{E}\|\mbox{POD}(a)-\mbox{POD}_{X_i}(a)\|^2}{D}\;,\\
T_i^{\mbox{\tiny POD}} & = & \displaystyle \frac{\mathbb{E}\|\mbox{POD}_X(a)-\mbox{POD}_{X_{-i}}(a)\|^2}{D}\;.
\end{array}
\end{equation}
These POD Sobol' indices are easily computed with the metamodels.
In particular, the kriging metamodel allows one to replace $\mathbb{P}( Y > s \mid a)$ by the expectation $\displaystyle \mathbb{E}_X \left[ 1 - \Phi\left(\frac{s-\widehat{Y_{P}}(a,X)}{\sigma_Y^2(a,X)}\right)\right]$ in the POD expressions of (\ref{eq:PODX}).

\begin{figure}[!ht]
  \centering
	\includegraphics[width=0.49\textwidth]{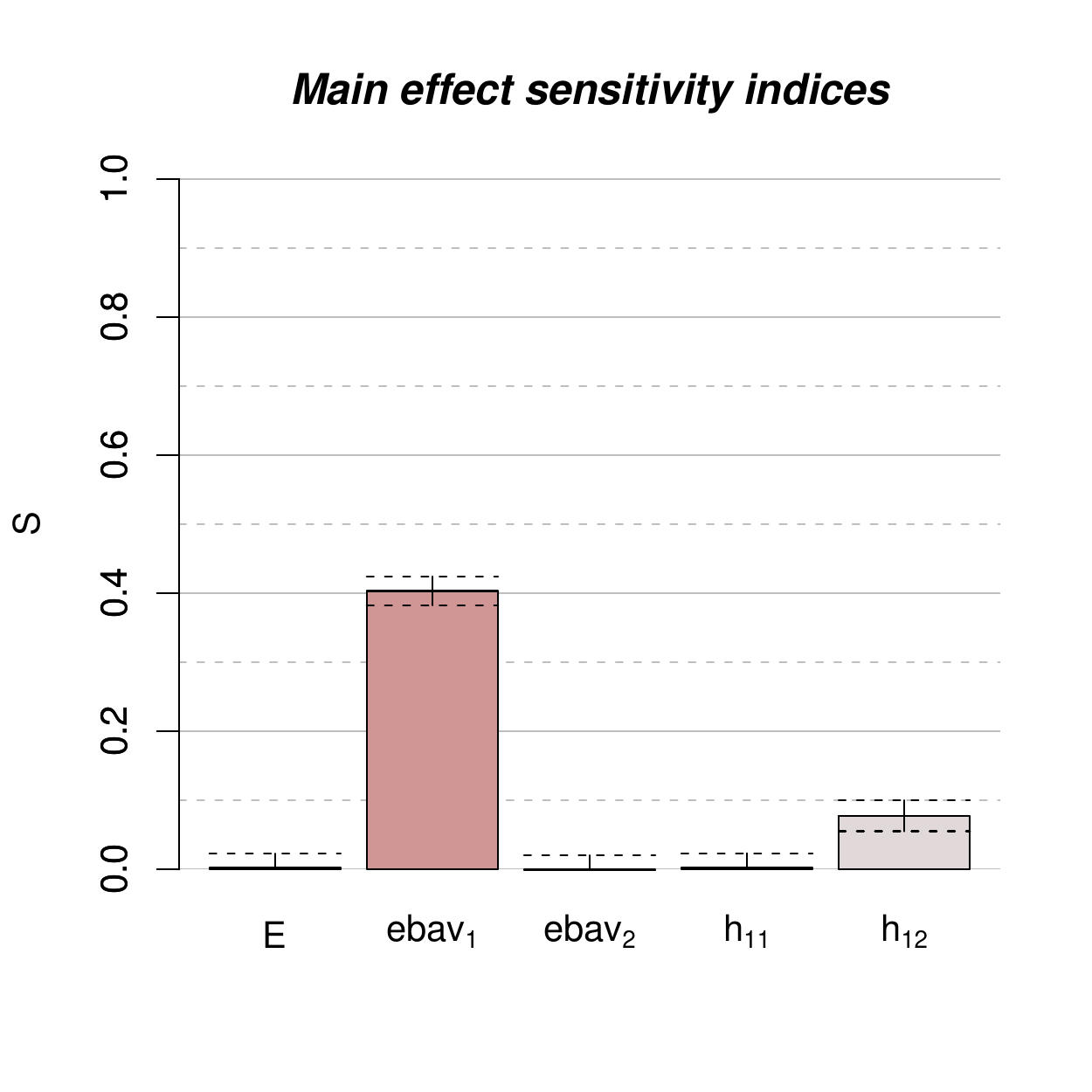}
	\includegraphics[width=0.49\textwidth]{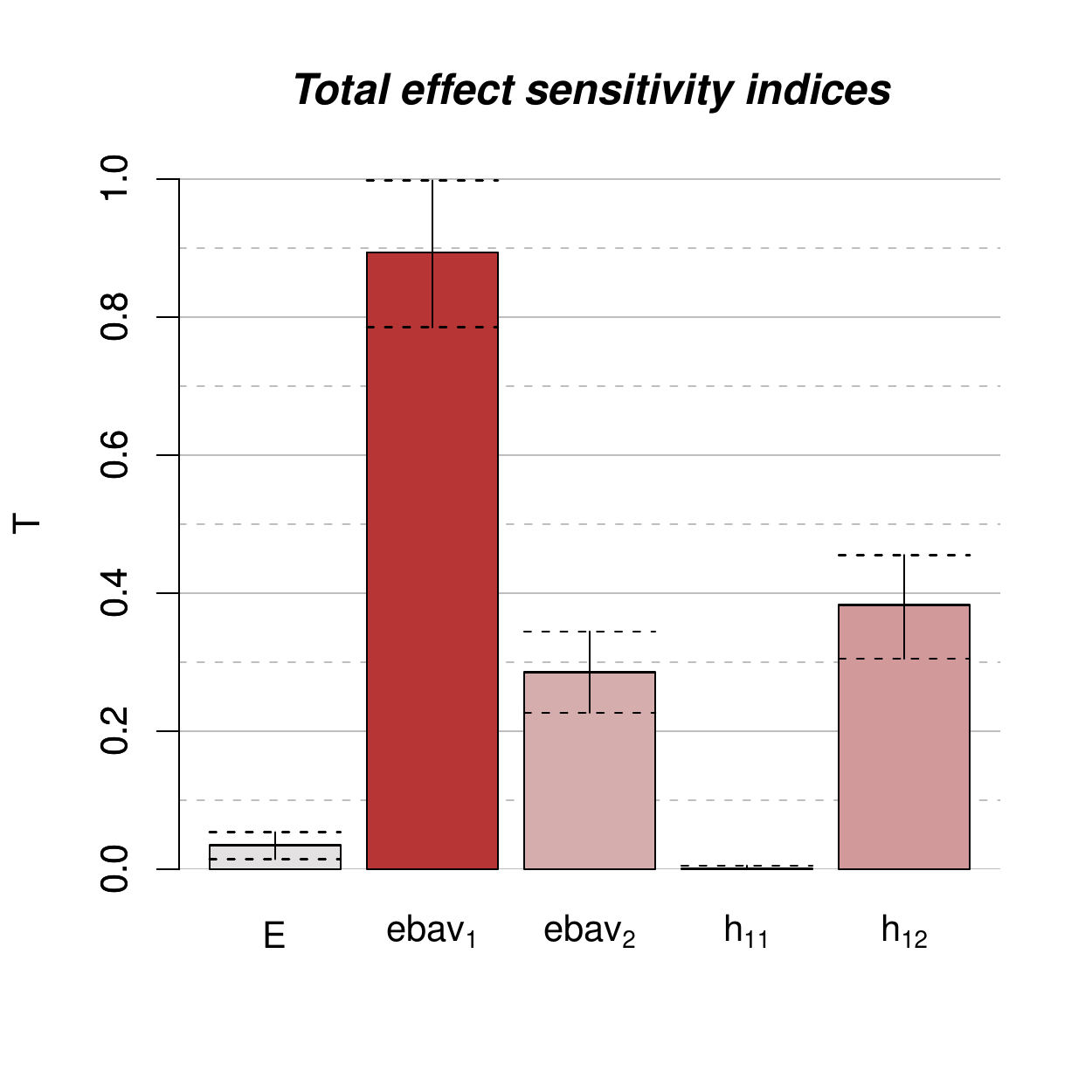}
\caption{POD first order (left) and POD total (right) Sobol' indices.}
  \label{fig:SobolPOD1}
\end{figure}

Figure \ref{fig:SobolPOD1} gives the sensitivity analysis results on our data.
We find that the POD curve is mainly influenced by $ebav_1$ parameter, with smaller effects of $ebav_2$ and $h_{12}$ parameters. 
As the first-order and total Sobol' indices strongly differ, we know that the main contributions come from interactions between these three influent parameters.
From an engineering point of view, working on the uncertainty reduction of $ebav_1$ is a priority in order to reduce the POD uncertainty.

\subsection{Sobol' indices for a specific defect size or probability}

The POD Sobol' indices quantify the sensitivity of each input on the overall POD curve.
However, we could be interested in the sensitivities on the detection probability at a specific defect size $a$.
As it is a scalar value, this can be directly done by replacing $Y$ by $\mbox{POD}_X(a)$ in all the equations of Section~\ref{sec:Sobol}.

\begin{figure}[!ht]
  \centering
	\includegraphics[width=0.49\textwidth]{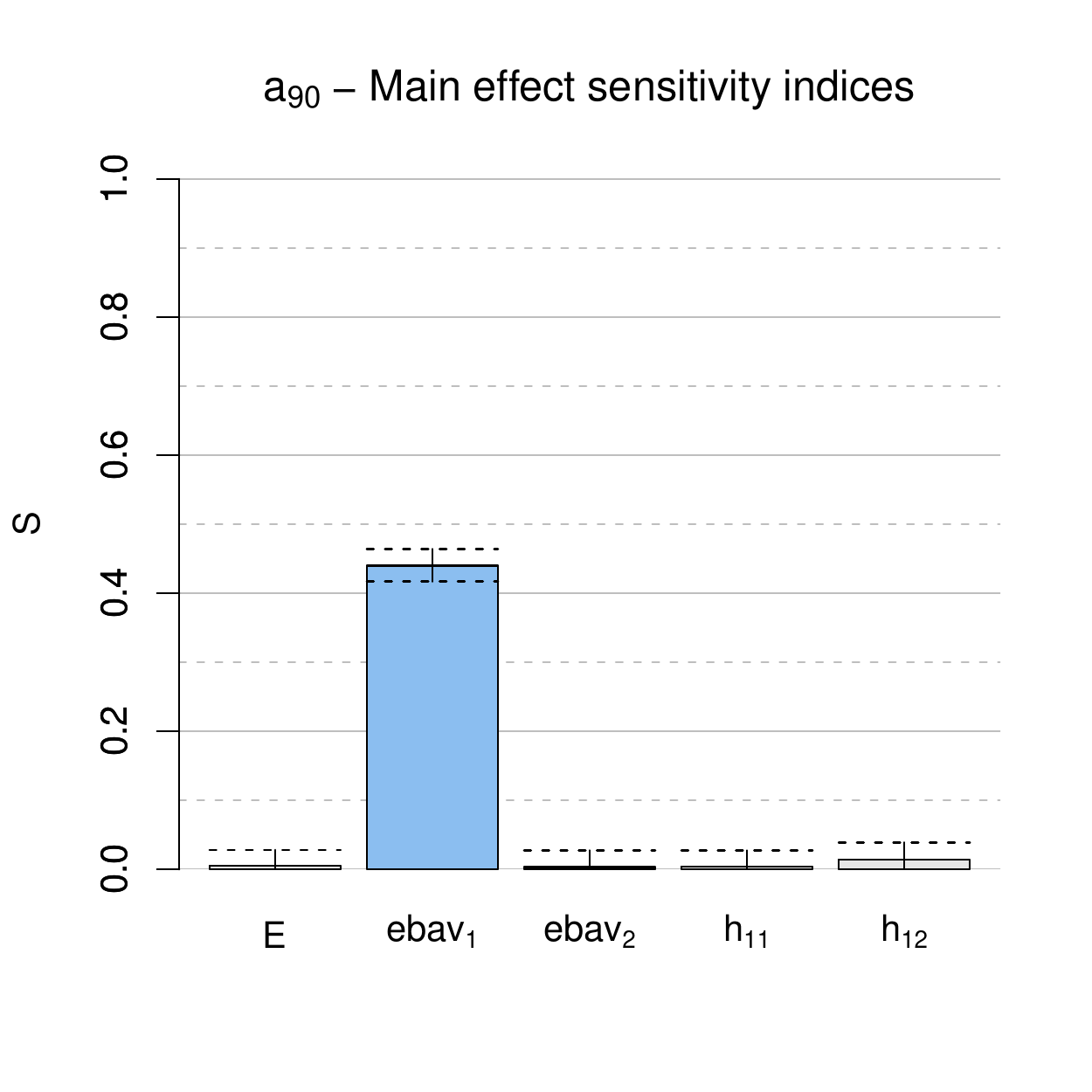}
	\includegraphics[width=0.49\textwidth]{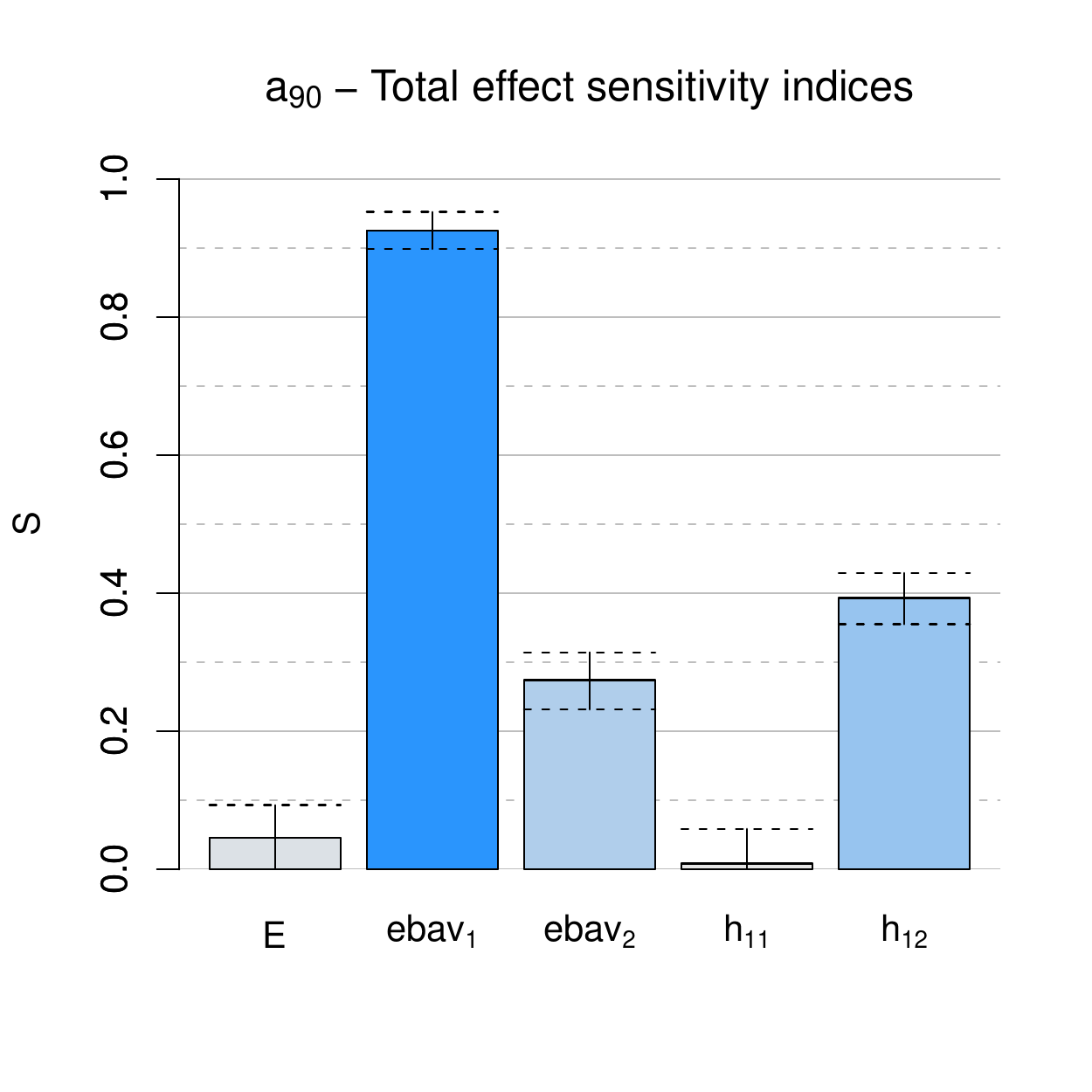}
\caption{First order (left) and total (right) Sobol' indices on $a_{90}$.}
  \label{fig:SobolPOD2}
\end{figure}

If we are now interested by the sensitivities on the defect size at a specific probability detection, we have to study the inverse function of the POD: $\mbox{POD}_X^{-1}(p)$ with $p$ a given probability.
Similarly to the previous case, the defect size Sobol' indices can be obtained by replacing $Y$ by $\mbox{POD}_X^{-1}(p)$ in all the equations of Section~\ref{sec:Sobol}.
Figure \ref{fig:SobolPOD2} displays these sensitivity indices on our data for $p=0.90$.
We conclude that $a_{90}$ is mainly influenced by $ebav_1$ parameter, with smaller effects of $ebav_2$ and $h_{12}$ parameters. 
The influences are similar than those of the POD curve.

\section{Conclusions}

This paper has presented four different techniques for POD curves determination (flaw detection probability), valuable over a wide range of NDT procedures.
As part of this study, we focus on the examination under wear anti-vibration bars of steam generator tubes with simulations performed by the finite-element computer code C3D.
The model parameterization and the design of numerical experiments have been firstly explained.

\begin{figure}[!ht]
  \centering
	\includegraphics[scale=0.5]{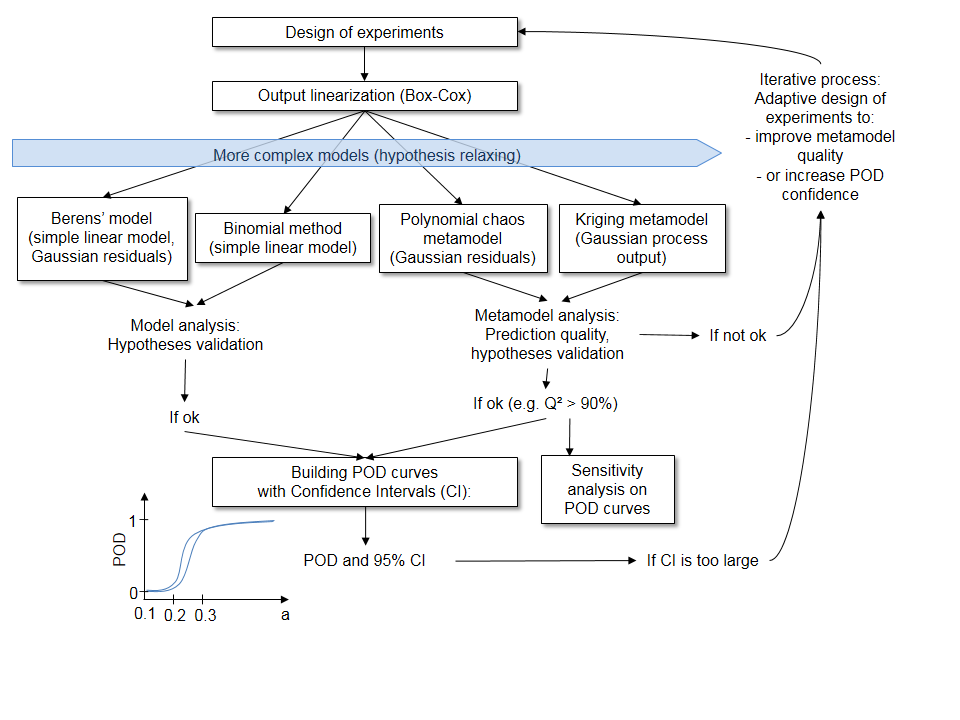}
\caption{General and progressive MAPOD methodology.}
  \label{fig:methodoPOD}
\end{figure}

Based on these methods of POD curves (and associated confidence intervals) determination, a general methodology is proposed in Figure \ref{fig:methodoPOD}.
It consists in a progressive application of the following methods:
\begin{enumerate}
\item the Berens method, based on a linear regression model, and requiring normality assumption on regression residuals;
\item the Binomial-Berens method which relaxes the normality hypothesis;
\item the polynomial chaos metamodel which does not require the linearity assumption but requires normal metamodel residuals;
\item the kriging metamodel.
\end{enumerate}
Other techniques, not discussed here, could be introduced in this scheme, as the quantile regression used in \citet{domfeu12} to relax Berens' hypothesis on the residuals distribution, or bootstrap-based alternatives.

The results of these four techniques in terms of the estimation of the defect size detectable with a $90\%$-probability ($a_{90}$) and its $95\%$-lower bound ($a_{90/95}$) are synthesized in Table \ref{tab:synt}.
While $a_{90}$ is rather unchanged, we observe slight variations on $a_{90/95}$ between the different methods.

\begin{table}[!ht]
$$\begin{tabular}{lcccc}
& Berens & Binomial-Berens & Polynomial chaos & Kriging\\ 
\hline
$a_{90}$ & 0.30 & 0.30 & 0.30 & 0.305 \\
$a_{90/95}$ & 0.31 & 0.305 & 0.32 & 0.315\\
\hline
\end{tabular}$$
\caption{Synthesis of results for detectable defect sizes (in mm) with the four methods of the POD methodology.}\label{tab:synt}
\end{table}

From the metamodel-based techniques, variance-based sensitivity analysis can also be performed in order to quantify the effect of each input on the POD curve.
Other sensitivity analysis methods devoted to POD curves allow to quantify the effects of the modifications of each input distribution.
For example, the Perturbation-Law based sensitivity Indices \citep{lemser13} would be the subject of a further work.
Finally, an iterative process can be applied to choose new simulation points in order to improve the metamodels predictivity or to reduce the POD confidence interval (see Fig. \ref{fig:methodoPOD}).
These metamodel-based sequential procedures have not been discussed in the present paper.

It is important to note that the obtained POD curves are based on a probabilistic modeling of system input parameters that has to be validated.
Moreover, the initial simple model (\ref{eq:model}) does not fully represent the reality, and taking into account the numerical model uncertainty is an important task \citep{aldkno13}.
Additional noise as reproducibility noise and measurement errors have also to be added.
Solutions for this problem, based on random POD models, are currently under study \citep{brofor15}.

\section{Acknowledgements}
Part of this work has been backed by French National Research Agency (ANR) through project ByPASS ANR-13-MONU-0011.
All the calculations were performed by using the OpenTURNS software~\citep{baudut16}.
We are grateful to L\'ea Maurice for initial works on this subject, as Pierre-Emile Lhuillier, Pierre Thomas, Fran\c{c}ois Billy, Pierre Calmon, Vincent Feuillard and Nabil Rachdi for helpful discussions.
Thanks to Dominique Thai-Van who provided a first version of Figure \ref{fig:methodoPOD}.

\bibliographystyle{spbasic}      
\bibliography{bib}   

\end{document}